\documentclass[a4paper]{article}

\usepackage{epsfig}

\usepackage{amsmath}
\usepackage{amssymb}
\usepackage{cite}

\usepackage{fancyhdr}
\numberwithin{equation}{section}
\numberwithin{figure}{section}

\newcommand{\biv}{P}            
\newcommand{\cas}{{\cal C}}

\newcommand{\beqs}{\begin{equation*}}
\newcommand{\beq}{\begin{equation}}

\newcommand{\eeqs}{\end{equation*}}
\newcommand{\eeq}{\end{equation}}

\newcommand{\beqas}{\begin{eqnarray*}}
\newcommand{\beqa}{\begin{eqnarray}}

\newcommand{\eeqas}{\end{eqnarray*}}
\newcommand{\eeqa}{\end{eqnarray}}

\newcommand{\eq}[2]{\begin{equation} #1 \label{#2} \end{equation}}

\newcommand{\meq}[2]{\begin{multline} #1 \label{#2} \end{multline}}

\newcommand{\eps}{\varepsilon}
\newcommand{\al}{\alpha}
\newcommand{\be}{\beta}
\newcommand{\ga}{\gamma}
\newcommand{\de}{\delta}
\newcommand{\om}{\omega}
\newcommand{\ka}{\kappa}

\newcommand{\blist}{\begin{itemize}}

\newcommand{\elist}{\end{itemize}}

\providecommand{\href}[2]{#2}

\newcommand{\twod}{$2D$}

\DeclareFontFamily{OT1}{rsfs}{}
\DeclareFontShape{OT1}{rsfs}{m}{n}{ <-7> rsfs5 <7-10> rsfs7 <10->rsfs10}{} 
\DeclareMathAlphabet{\mycal}{OT1}{rsfs}{m}{n}
\newcommand{\scri}{{\mycal I}}

\begin{document}


\begin{titlepage}

\renewcommand{\thefootnote}{\fnsymbol{footnote}}

\hfill TUW--03--16

\hfill LU-ITP 2004/008

\hfill ESI 1461

\begin{center}

\vspace{0.5cm}

{\Large\bf Long time black hole evaporation with bounded Hawking flux} 

\vspace{1.0cm}

{\bf D.\ Grumiller\footnotemark[1]}

\vspace{5ex}

  {\footnotemark[1]\footnotesize Institut f\"ur
    Theoretische Physik, Technische Universit\"at Wien \\ Wiedner
    Hauptstr.  8--10, A-1040 Wien, Austria
\vspace{2ex}

and

\vspace{2ex}

Institut f\"ur Theoretische Physik, Universit\"at Leipzig \\
Augustusplatz 10-11, D-04103 Leipzig, Germany}

\vspace{2ex}

   \footnotetext[1]{e-mail: \texttt{grumil@hep.itp.tuwien.ac.at}}
\end{center}
\vspace{5ex}

\begin{abstract}


The long time behavior of an evaporating black hole presents a challenge to theoretical physics and touches relevant conceptual issues of quantum gravity, such as the information paradox. There are basically two strategies: top-down, i.e., to construct first a full quantum theory of gravity and to discuss black hole evaporation as a particular application thereof, and bottom-up, i.e., to sidestep the difficulties inherent to the former approach by invoking ``reasonable'' ad-hoc assumptions. 

Exploiting the fact that the Schwarzschild black hole can be described by means of an effective theory in $2D$, a particular dilaton gravity model, the latter route is pursued. A crucial technical ingredient is Izawa's result on consistent deformations of $2D$ BF theory, while the most relevant physical assumption is boundedness of the asymptotic matter flux during the whole evaporation process. Together with technical assumptions which can be relaxed, the dynamics of the evaporating black hole is described by means of consistent deformations of the underlying gauge symmetries with only one important deformation parameter. An attractor solution, the endpoint of the evaporation process, is found. Its metric is flat. However, the behavior of the dilaton field (which corresponds to the surface area) is nontrivial: it is argued that during the final flicker a first order phase transition occurs from a linear to a constant dilaton vacuum. Consequently, a shock wave is emitted as a final ``thunderbolt'' with a total energy of a fraction of the Planck mass. Relations to ultrarelativistic boosts are pointed out. Another fraction of the Planck mass may reside in a cold remnant. 

The physical discussion addresses the life time, the specific heat, the Carter-Penrose diagram, the information paradox and cosmological implications. The phenomenon of ``dilaton evaporation'' to a constant dilaton vacuum might be of relevance also for higherdimensional scalar tensor theories.


\end{abstract}


\vfill
\end{titlepage}



\section{Introduction}\label{se:1}

For more than three decades the physics of black holes (BHs) has attracted the attention of an increasing number of relativists, astrophysicists, string theoreticians and elementary particle physicists (for a recent textbook cf.\ e.g.\ \cite{Frolov:1998}). One of the most spectacular theoretical predictions is the Hawking effect: due to quantum creation of particles a BH may evaporate with ensuing radiation having a thermal distribution \cite{Hawking:1975sw}.

However, there are well-known problems with the long-time evolution of BHs (e.g.\ in the form of the information paradox---cf.\ e.g.\ \cite{Banks:1995ph})). Although semi-classical models have established interesting insights, especially in the context of twodimensional (henceforth $2D$) dilaton gravity (cf.\ e.g.\ \cite{Thorlacius:1995ip} with particular focus on the model by Russo, Susskind and Thorlacius) 
the main obstacle in such a discussion stems from the fact that a comprehensive theory of quantum gravity does not yet exist.
Moreover, already simple classical systems of gravity with matter are not exactly soluble which, in general, prohibits a nonperturbative treatment of quantum backreactions. 

The present approach tries to circumvent these difficulties at the cost of several ad-hoc assumptions, some of which could be called ``natural''. The main idea is that one allows for a continuous family of models labelled by one or more deformation parameters. The word ``deformation'' will be given a precise meaning when the assumptions are specified, but roughly speaking one has a family of actions labelled by these parameters. The difficult task is to establish an evolution equation for the latter. Dynamics then will dictate the evolution of these parameters. If an attractor solution exists dynamical deformations imply the transition of the original model, e.g. the Schwarzschild black hole (SS BH), to a specific remnant geometry, namely the attractor. Thus, rather than trying to find one model describing an evaporating BH from the early stages until the very end we have a family of models at our disposal and at each instant one family member provides the most adequate description of geometry. Provided the notion of light-like infinity ($\scri$) makes sense---this will be one of the ad hoc assumptions---one can imagine an asymptotic observer who patches all these models together along $u=\rm const.$ lines, where $u$ is the retarded time.

To be more concrete, the assumptions chosen in the present paper are 1.\ the restriction to consistent deformations in the technical sense of Barnich and Henneaux (roughly speaking, neither gauge degrees of freedom nor physical degrees of freedom may appear or disappear, but the gauge symmetries may be deformed, e.g.\ from an abelian gauge symmetry to a nonabelian one), 2.\ the assumption of asymptotic flatness of each spacetime solving the equations of motion of the deformed action during the whole process of evaporation, 3.\ the absence of nonextremal horizons to avoid extremal endpoints of the evolution (actually, a further technical assumption will be imposed which, however, can be relaxed), 4.\ an observer at spatial infinity ($i^0$) is assumed who measures a constant Hawking flux, i.e.\ the increase of energy loss of the black hole due to the Hawking effect is compensated by a corresponding boost of the observer. It is emphasized that this is merely a technical trick to allow for an easier derivation of the attractor solution. Of course, consequently practically nothing can be said about what happens at a given instant during the evaporation in terms of ``physical time''---for this information not only the existence of such a boost but the precise value of the boost parameter would be needed. But since our task is much more humble---we are only interested in the specific form of the remnant geometry and not a thorough description of the complete dynamics---these assumptions will be sufficient to provide a unique answer for the remnant geometry.

The paper is organized as follows: the four assumptions relevant for the SS BH are presented in sect.\ \ref{sec:3}. As a result due to the evaporation process the SS BH is deformed towards an effectively $2+1$ dimensional model---in more picturesque (albeit slightly misleading) terms the celestial sphere evaporates into a celestial circle. A physical discussion of the results (e.g.\ the life time of a SS BH, its line element, the relevance of Planck scale contributions, dilaton evaporation, consequences for the information paradox and cosmological implications) in sect.\ \ref{sec:4} and possible relaxations of the assumptions (sect.\ \ref{sec:5}) conclude this work. Complementary material can be found in the appendices: appendix \ref{app:def} recapitulates deformations of generic dilaton gravity; a possible generalization which relaxes the technical assumptions is discussed in appendix \ref{app:A}. A simple toy model from classical mechanics which mimics some of the main features of BH evaporation as described in the present work can be found in appendix \ref{sec:cm}.

\section{Evaporation of a Schwarzschild black hole}\label{sec:3}

The SS BH is the simplest of phenomenologically relevant BH models. Once the long time behavior of this case is understood it appears that generalizations to Reissner-Nordstr\"om or Kerr-Newman are comparably easy. More will be said on this in sect.\ \ref{sec:5}; for now the focus will be on the SS BH. The formation process is completely neglected in these considerations. It is assumed that space-time has ``settled down'' after the collapse of matter and an approximately stationary situation is encountered.

A ``natural'' working hypothesis, namely that no deformation occurs and spacetime remains SS for all times, leads to the usual prediction of an explosive evaporation during the final stage, being the result of applying the semi-classical approximation beyond its limit of validity, because eventually quantum backreactions will deform geometry appreciably. 

It is noteworthy that no assumption is made regarding the type of matter coupled to gravity. It is just taken for granted that some matter degrees of freedom supporting the Hawking radiation actually do exist. Specifying matter explicitly has the disadvantage that one would have to find an exact solution of the coupled matter-gravity system in order to study the long time evolution of an evaporating BH. The present approach circumvents these difficulties at the price of several ad hoc assumptions.\footnote{It should be pointed out that the idea to sidestep the difficulties of quantum gravity and nevertheless obtain relevant predictions clearly is not new and has been pursued, for instance, in the framework of Doubly Special Relativity (for a recent review and more references cf.\ e.g.\ \cite{Amelino-Camelia:2000mn}) 
or in the context of $\ka$-deformations \cite{Lukierski:1991pn}. 
Recent generalizations of some of these results to ``gravity with an invariant energy scale'' exist \cite{Mignemi:2002hd}, 
but one has to be careful with the definition of the line-element \cite{Grumiller:2003df,Strobl:2003kb}. A different philosophy which appears to be closer to the one of the present work consists of taking the Hilbert-Einstein action without further structure as starting point \cite{Bonanno:2000ep}. The key ingredient in that reference is a running Newton constant. The causal structure is like that of a Reissner-Nordstr\"om BH because an inner horizon develops. The outer horizon decreases, like in the present approach; there seems to be a regular dS core (as suggested by \cite{Frolov:1988vj}). However, as a careful analysis in their work shows nonetheless a singularity at $r=0$ is present, because the Newton constant scales differently in dS and SS near $r=0$. A minimal BH size of order of Planck mass is found and a cold remnant is reached after infinite time (note: a similar cold remnant has been found also in \cite{Alexeyev:2002tg} for Einstein-Dilaton-Gauss-Bonnet gravity).} 

\subsection{The four assumptions}\label{sec:2}

In many instances---take for example the singularity theorems in General Re\-la\-ti\-vi\-ty---it helped to distinguish between the principal part of assumptions (e.g.\ one has to specify an energy condition) and their specific form (e.g.\ the strong energy condition). Similarly, one can relax or generalize the assumptions presented below. Some of these generalizations are addressed in sect.\ \ref{sec:5} and appendix \ref{app:A}. On a first reading it might be useful to consult appendix \ref{sec:cm} at this point.

For each of the four assumptions first its realization in the toy example in appendix \ref{sec:cm} will be mentioned, then the general form will be presented and finally its specification relevant to the core part of this work together with its implementation.

\subsubsection{First assumption: Consistent deformations}

\paragraph{Toy example:} Due to interactions with the unknown ``matter'' part the geometric system (\ref{eq:rev1}) can be deformed by consistent deformations.

\paragraph{General form:} Due to quantum gravity effects during the long-time evolution of a BH, the gauge symmetries of gravity may be deformed in a {\em specific way} imposed by hand (e.g.\ allow no deformations at all or only a certain class thereof). In principle, there can be an arbitrary amount of deformation parameters, but of course it is advantageous to keep its number as low as possible.

\paragraph{Specific form:} The BH evaporation due to the Hawking effect (plus eventual non-per\-tur\-ba\-tive effects from quantum backreactions) does not change the number of gauge degrees of freedom or physical degrees of freedom---i.e.\ the evaporation induces only consistent deformations\footnote{Because of the rigidity result for consistent deformations of Einstein(-Yang-Mills) theory of gravity in $D>2$ \cite{Barnich:1995ap} (at least) one of its premises must be violated if nontrivial modifications of the gauge symmetries are required to arise. In the present paper it is the dimension, $D=2$, and the nonEinsteinian nature of the (dilaton deformed) nonlinear gauge symmetries. \label{fn:7}} of the underlying model (in the sense of Barnich and Henneaux \cite{Barnich:1993vg}). Since spherical reduction and consistent deformations will not commute in general, it should be pointed out that consistent deformations are applied solely to the reduced theory. Thus, we are not dealing with the SS BH in the framework of Einstein gravity in $D=4$, but with its (classically equivalent) counterpart as $2D$ dilaton gravity. Obviously, deformations exciting nonspherical modes cannot be treated in such a simple manner. This situation resembles the so-called ``spherical reduction anomaly'': spherical reduction and renormalization do not commute \cite{Frolov:1999an}. 

\paragraph{Remarks:} Because of quantum backreaction effects, which are taken into account implicitly, the deformation parameters are not necessarily constant during the evaporation process but in general some function (e.g.\ of the retarded time $u$). The first postulate implies that the action may change continuously with the retarded time $u$ such that effectively one has to patch the classical solutions of different models at lines $u=\rm const$. It is not intended to construct one geometric model describing the BH evolution from the early stages until the final flicker. Rather, we have a continuous family of models at our disposal, labelled by the deformation parameters, and at each stage of the process one of them is the most adequate to describe geometry. 

\paragraph{Implementation:} The crucial first condition restricts the possible deformations of classical gravity due to quantum effects to consistent deformations. Classically, the SS BH can be derived from a Poisson-sigma model (PSM) with three target space coordinates \cite{Schaller:1994es}, one of which can be identified with the so-called dilaton field (for the SS BH the dilaton field is proportional to the ``surface area''). Of course, classical equivalence by no means implies quantum equivalence, but nevertheless the PSM formulation will be taken as starting point. Izawa has shown that the most general consistent deformation of a PSM is another PSM with the same number of target space coordinates \cite{Izawa:1999ib}. Thus, the first assumption restricts to the class of PSMs with three target space coordinates. It contains infinitely many possible models, but with a little bit of extra structure each of them can be interpreted as a $2D$ dilaton gravity model \cite{Strobl:2003kb}. For most discussions this first order formalism, which uses the Cartan variables as basic fields, is superior---for a comprehensive review cf. e.g.\ \cite{Grumiller:2002nm}. However, because more people are familiar with the second order formalism and since few technical details are needed the first order version of (\ref{vbh:sog}) will not be employed, apart from the self-contained concise presentation in appendix \ref{app:def}. The second order action reads \cite{Odintsov:1991qu} 
\begin{equation}
 L^{\rm SOG}=\int_{\mathcal{M}_2} d^2 x \sqrt{-g} \left[
X\frac R2 -\frac{U(X)}2 (\nabla X)^2 +V(X) \right]\,, 
\label{vbh:sog}
\end{equation}
where $X$ is the dilaton, $g$ the $2D$ metric and $R$ the Ricci scalar. The potentials $U,V$ are arbitrary functions establishing the ``$U-V$ family of models'' (cf.\ appendix \ref{app:A}). It covers practically all relevant dilaton gravity models, including the CGHS \cite{Callan:1992rs}, the Jackiw-Teitelboim model \cite{Barbashov:1979bm}, 
the Katanaev-Volovich model \cite{Katanaev:1986wk}, 
and spherically reduced gravity \cite{Berger:1972pg}. 
By deforming the model in principle one can leave this class, but for simplicity it will be assumed that this does not happen.  

\subsubsection{Second assumption: Asymptotic flatness}

\paragraph{Toy example:} The ``asymptotic'' region hosting the observer is fixed by the asymptotic condition $q(0)=0$.

\paragraph{General form:} The asymptotics of space-time has to be fixed in some way---typically by requiring asymptotic flatness during the whole evaporation process, but different behavior (such as asymptotic (A)dS) is conceivable.

\paragraph{Specific form:} The spacetime is assumed to be asymptotically flat during the whole process of evaporation (``stability of $i^0$ and $\scri$'').

\paragraph{Remarks:} This corresponds to a far field approximation---it is assumed that the observer measuring the Hawking flux is sufficiently far away from the BH and that this feature does not change during the evaporation process. It implies that at least for a distant observer the continuous patching of models is sensible, since all models have essentially the same asymptotics by construction. 

\paragraph{Implementation:} Within the PSMs this assumption constrains us to the important class of Minkowski ground state (MGS) theories, i.e.~to those models which contain Minkowski space among their classical solutions. The MGS condition 
implies in the context of the $a-b$ family defined below the linear relation $a=b+1$. 
 
\subsubsection{Third assumption: Avoidance of extremality}

\paragraph{Toy example:}  There are restrictions on the ``causal structure'': the deformed potential in (\ref{eq:rev2}) is assumed to depend only on $q$ and it must not have an extremum besides $q=0$. As a technical simplification $V\propto q^2$ has been required, but supposedly this can be relaxed without changing the attractor solution.

\paragraph{General form:} Restriction of the causal structure (e.g.\ by requiring that the number of (non-extremal) horizons is non-increasing).

\paragraph{Specific form:} At most one (non-extremal) apparent horizon is present. For technical reasons only deformations within the so-called $a-b$ family of dilaton gravity models will be considered \cite{Katanaev:1997ni}, i.e.\ potentials of the form
\eq{
U(X)=-\frac{a}{X}\,,\quad V(X)=-\frac{B}{2}X^{a+b}\,,\quad a,b,B\in\mathbb{R}\,,
}{3.64}
to be inserted into the action (\ref{vbh:sog}). All of them exhibit at most one (non-extremal) horizon.
 
\paragraph{Remarks:} This is the most technical and from a general point of view the least essential assumption; it could be that it can be dropped in certain generalizations, but for safety it is included here.  It is needed to avoid extremal cases, both in the literal and the transferred sense. It can be considered as a consequence of the Penrose theorem (an isolated BH cannot split into two or more BHs, cf.\ e.g\ \cite{Hawking:1973}), although of course one has to be careful\footnote{E.g.\ in the Frolov-Vilkovisky model \cite{Frolov:1981mz} although classically only one horizon exists at the quantum level an inner (apparent) horizon emerges. I am grateful to V.\ Frolov for enlightening discussions on this model.} by applying these classical theorems to BH evaporation (clearly Hawking's area theorem \cite{Hawking:1971tu} 
is violated).

\paragraph{Implementation:} In this simplified scenario the only deformation parameters are $a,B\in\mathbb{R}$. For SS $a=1/2$. The parameter $B$ is almost irrelevant as it just defines the physical scale. Note that $a,b$ appear nonlinearly while $B$ enters only linearly. The generalization to the $U-V$ family is treated in appendix \ref{app:A}.

\subsubsection{Fourth Assumption: Boundedness of Hawking flux}

\paragraph{Toy example:} The ``toy model Hawking temperature'' (\ref{eq:rev5}) remains constant during the whole dynamical process (actually this can be replaced by the much weaker condition of boundedness and positivity by a reparametrization of the asymptotic observer).

\paragraph{General form:} The fourth assumption implicitly supposes that the notion of an asymptotic Hawking flux does make sense. While this is definitely true for the early stages of the evaporation it has to be regarded as a working hypothesis for the final evolution. Because this can be considered as controversial, before presenting the actual assumption some preemptive comments are in order: as mentioned above, we are not dealing with one model, but with a whole family. Each family member describes e.g.\ an eternal BH and as such allows for a sensible definition of Hawking flux and temperature. It is emphasized that at each instant the semi-classical approximation implicit in the notion of Hawking flux has to be fulfilled only for an infinitesimal amount of time. The crucial assumption is boundedness and positivity (or some stronger requirement) of the Hawking flux as measured by an asymptotic observer.

\paragraph{Specific form:} An observer at $i^0$ is assumed to be isothermal during the whole evaporation process, i.e.\ the asymptotic Hawking flux does not change with the retarded time $u$ (apart from the points where it is turned on and off). 

\paragraph{Remarks:}  Physically, this is the most important assumption because boundedness typically implies that the ensuing dynamical system to be derived from these assumptions encodes effects beyond the semi-classical approximation and positivity ensures that no exotic matter flux is measured in the asymptotic region. The requirement of boundedness of the Hawking flux is in the same spirit as the so-called Limiting Curvature Hypothesis (cf.\ \cite{Trodden:1993dm} for an application to {\twod} BHs) which postulates that due to quantum gravity effects curvature remains bounded (thus implicitly assuming that the notion of curvature still makes sense). In that context one has to fix the potentials $U,V$ in (\ref{vbh:sog}) by hand appropriately. In this manner an eternally radiating remnant has been predicted \cite{Easson:2002tg}. In the rest of this paper the terms ``asymptotic Hawking flux'' and ``Hawking temperature'' will be used interchangeably.

\begin{figure}
\setlength{\unitlength}{0.01\linewidth}
\begin{picture}(100,30)(0,10)
\put(0,10){\mbox{\resizebox{0.4\linewidth}{0.24\linewidth}{\includegraphics{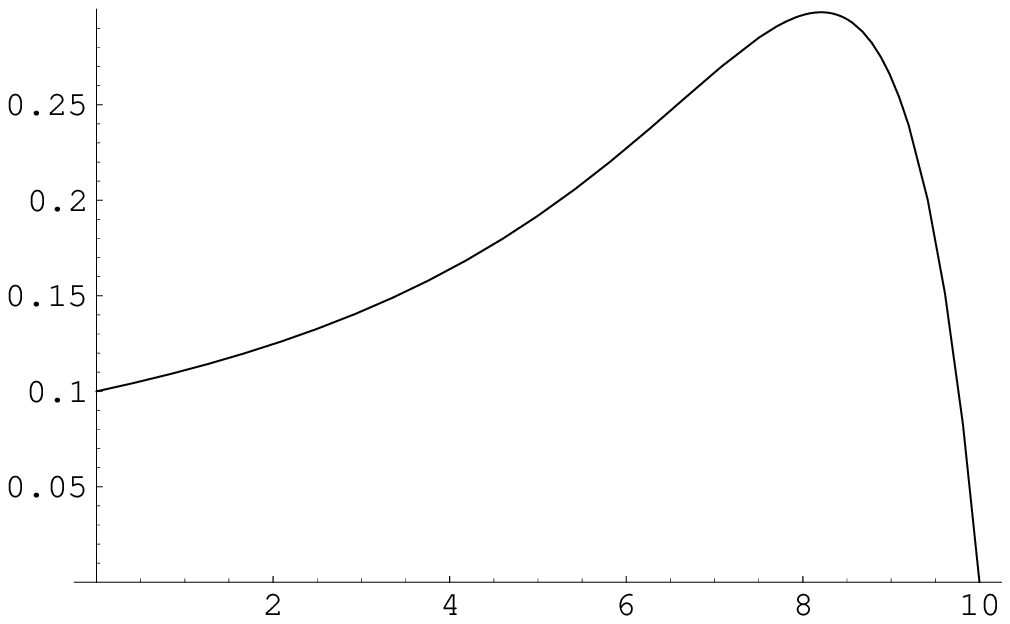}}}}
\put(1,37){\makebox(0,0)[l]{$\tilde{T}_H$}}
\put(42,12){\makebox(0,0)[l]{$\tilde{u}$}}
\put(45,23){\makebox(0,0)[l]{$\longrightarrow$}}
\put(55,10){\mbox{\resizebox{0.4\linewidth}{0.24\linewidth}{\includegraphics{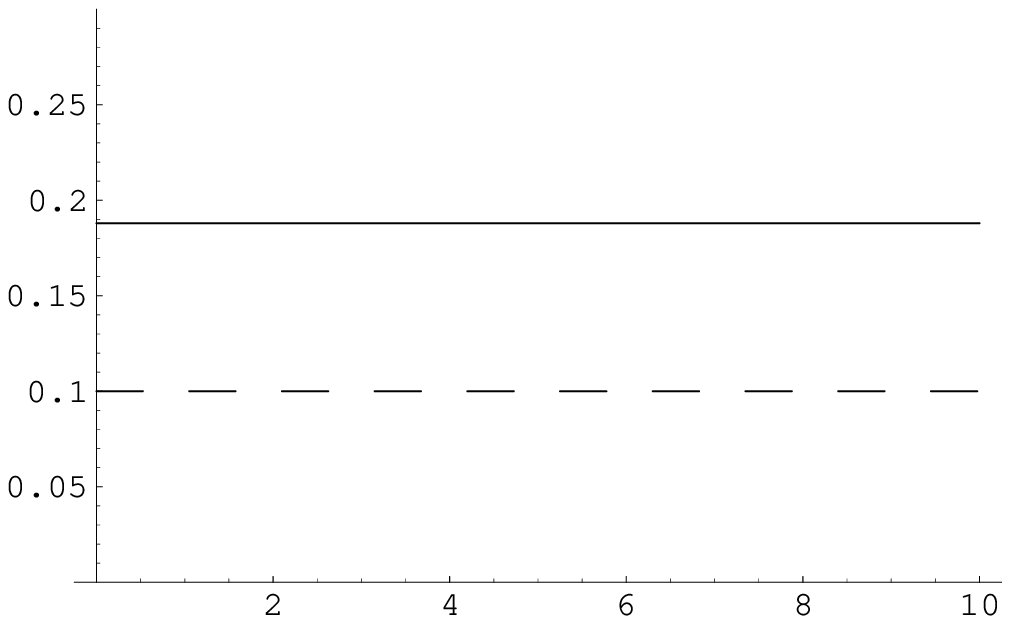}}}}
\put(56,37){\makebox(0,0)[l]{$T_H$}}
\put(97,12){\makebox(0,0)[l]{$u$}}
\end{picture}
\caption{Example of a transformation to constant $T_H$}
\label{fig:dif}
\end{figure}

\paragraph{Implementation:} The fourth assumption is probably the least trivial one. In fact, it can be replaced by the much weaker condition that the Hawking temperature $\tilde{T}_H$ remains bounded and non-vanishing in a finite interval at $\scri^+$. This is not unplausible from a physical point of view: it is supposed that quantum backreaction effects prevent the unbounded growth of the asymptotic flux as predicted by semi-classical calculations; moreover, the Hawking temperature does not vanish up to the point where the evaporation terminates. The asymptotic observer will be able to measure a temperature profile $\tilde{T}_H(\tilde{u})$ as a function of the retarded time $\tilde{u}$. Imposing that the Hawking process starts at $\tilde{u}=0=u$ by a regular diffeomorphism
\eq{
u(\tilde{u})=\frac{1}{T_H}\int_0^{\tilde{u}}\tilde{T}_H(\tilde{u}')d\tilde{u}'
}{utrafo}
the new profile $T_H(u)$ is constant (see fig.\ \ref{fig:dif}). 
It is convenient to rescale $T_H$ linearly such that it coincides with $\tilde{T}_H(\tilde{u}=0)$ because in this way $u$ and $\tilde{u}$ are practically identical during the early stages of the evaporation process (dashed line in fig.\ \ref{fig:dif}). Some physical interpretation relevant for the noninertial observer is presented in sect.\ \ref{sec:new}. We recall that the evolution is understood to be within a continuous family of models, and not within a single model; thus, the diffeomorphism (\ref{utrafo}) is not a diffeomorphism acting on a single metric, but rather redefines the way in which the models are glued together. The knowledge of the ``correct'' profile $\tilde{T}_H(\tilde{u})$ would be needed for determination of the precise form of the function $f(u)$ in the Bondi mass, eq.\ (\ref{iso16}) below. While the knowledge of $f(u)$ would be necessary to answer questions like ``how does the BH look like $10^x$ years after the evaporation starts?'' the goal of the present analysis will be much more moderate: we are only able to determine the possible endstates of BH evolution. Thus, we are asking the following question instead: ``With the given assumptions leading to a well-defined family of possible models and a well-defined evolution within this family, which are the attractor solutions?'' As will be shown below it can be answered by exploiting isothermality $T_H=\rm const.$

\subsection{Exploiting isothermality}

For the $a-b$ family of models obeying the MGS condition the Hawking temperature $T_H$ reads\footnote{Natural units $c=\hbar=G_N=k_B=1$ will be used exclusively. Thus, $M_{\rm Planck}=1$.} \cite{Liebl:1997ti}
\eq{
T_H=f(B,a)(2M_{BH})^{(a-1)/a}\,,
}{iso0}
where $M_{BH}$ is the BH mass, $a$ is the family parameter and $f(B,a)$ is a proportionality factor. $T_H$ is related to the asymptotic Hawking flux $T_{\rm flux}^{\rm asy}$ via the Stefan-Boltzmann law in $2D$:
\eq{
T_{\rm flux}^{\rm asy}=\frac{\pi}{6}T_H^2\,.
}{iso0.5}
At each instant the system behaves as if it were an eternal BH with a mass-to-temperature law as given by (\ref{iso0}) in accordance with the discussion in sect.\ \ref{sec:2}. Infinitesimal changes of the temperature can be induced by changes of the black hole mass, changes of the family parameter (deformations) and changes of the normalization constant $f$:
\eq{
\frac{dT_H}{T_H}=\left[\frac{dM_{BH}}{M_{BH}\ln{(2M_{BH})}}-\frac{da}{a(1-a)}\right] \frac{a-1}{a}\ln{(2M_{BH})} + \frac{df}{f}
}{iso4}
It turns out that the first terms dominate over the Planck scale contribution provided that $M_{BH}$ is large enough. The term $df/f$ is called ``Planck scale contribution'' because of the identity $df=dT_H(M_{BH}=1/2)+2f\frac{1-a}{a}dM_{BH}(M_{BH}=1/2)$. For large initial BH masses it can be neglected because the relevant terms in (\ref{iso4}) scale with $\ln{M_{BH}}$ while the $df$-term is of order of unity. This is the reason why the scale parameter $B$ is irrelevant until the final stage. For the moment this term will be dropped, but it will be reconsidered in in sect.\ \ref{se:3.4}. In the rigid case $df=0=da$ the well-known relation (cf.~e.g.~eq.~(6.29) of \cite{Grumiller:2002nm} and ref.~\cite{Liebl:1997ti})
\eq{
\frac{dT_H}{T_H}=\frac{a-1}{a}\frac{dM_{BH}}{M_{BH}}\,,
}{iso5}
is recovered (the choice $a=1/2$ for the SS BH yields the famous inverse proportionality between mass and temperature).

Isothermality $dT_H\stackrel{!}{=}0$ implies (for $a\neq 0$, $a\neq 1$, $df=0$ and $M_{BH}\neq 1/2$) an ordinary differential equation which can be integrated from $M_{i}$ to $M_{f}$ and from $a_i$ to $a_f$, establishing
\eq{
x_f=x_i\,z\,,\quad z:=\frac{\ln{(2M_f)}}{\ln{(2M_i)}}\,,\quad x_{i,f}:=\frac{a_{i,f}}{1-a_{i,f}}\,,
}{iso2}
which is already the main result of these considerations. 

\subsection{The CGHS model as repellor}

Applying the previous analysis to a given starting model (e.g.\ $a_i=1/2$ for SS) with a given initial BH mass $M_i\gg 1$ one can predict with (\ref{iso2}) the ``final'' model at a given smaller BH mass. For $z\approx 1$ one obtains $a_f\approx a_i$, as expected. Due to the logarithmic behavior of $z$ the BH has to radiate most of its mass before the model is deformed appreciably (for instance $z=1/2$ is induced by $2M_f=\sqrt{2M_i}$, which is just a tiny fraction of the original mass, but still large as compared to $M_{\rm Planck}$). The main questions is: what happens to $a_f$ for monotonically decreasing $z$ (especially in the limit $z\to 0$)? It can be answered straightforwardly: the parameter $a_f$ tends to zero unless $a_i\ge 1$ (or $a_i=\pm\infty$). Note that $z=0$ corresponds to $M_f=M_{\rm Planck}/2$. It will be assumed for the moment that this is the endpoint of the evaporation process.  

In language of dynamical systems the point $a=0$ is an attractor (a fixed point of the evolution which attracts all initial values which are close enough---in the present case close enough means $a_i\in(-\infty,1)$), while the CGHS model is a repellor\footnote{It should be pointed out that for BH masses smaller than $M_{\rm Planck}/2$ repellor and attractor change their roles: in this regime almost all models are driven towards the CGHS. But in this regime the contribution from $df$ no longer is negligible (see next subsection).} (a fixed point of the evolution which repels all other initial values): each model with $a=1-\eps$, $\eps>0$ is driven towards the attractor, while each model with $a=1+\eps$, $\eps>0$ leads to a runaway solution $a_f\to\infty$ because for small enough values of $z$ a pole is reached in (\ref{iso2}). So $a_f=\infty$ is an asymptotic fixed point which attracts all initial values $a_i\in(1,\infty)$. The other asymptotic fixed point $a_f=-\infty$ is unstable.

\subsection{Evolution close to the attractor}\label{se:3.4}

Before addressing the further evolution of the attractor solution it is worthwhile to discuss it geometrically. At the $2D$ level the gauge symmetries of the model are related to a nonlinear (finite W) algebra and thus a geometric discussion in simpler terms would be desirable. To this end it is helpful to note that all models on the line $a=b+1$ correspond to spherically reduced theories, where the dimension of the isometry sphere is given by $D=1/(1-a)$ (so for $a=1/2$ the 2-sphere corresponding to the isometry orbits of SS is obtained). In a sense, Hawking radiation induces an evaporation of all of the dimensions of the sphere but one, but the limiting solution in general does not correspond to ``spherically'' reduced Einstein gravity from $2+1$ dimensions. Only for $B=0$ the model corresponds to a toric reduction of $2+1$ dimensional Einstein gravity with the trivial line element $(ds)^2=2drdu+(du)^2-r^2(d\phi)^2$. Note that it is not necessary to adopt the viewpoint of ``evaporating dimensions''---alternatively, one can adopt a purely $2D$ point of view. It just seems easier to interpret the deformed gauge symmetries in terms of spherically reduced Einstein theory, because most people are accustomed to the gauge symmetries of the latter.

The $2D$ line element (in Eddington-Finkelstein gauge) reads
\eq{
(ds)^2=2drdu+\left(B\ln{r}-2M_{BH}\right)(du)^2\,.
}{iso14}
The curvature scalar $R=-2B/r^2$ vanishes for $r\to\infty$. The conformal diagram for (\ref{iso14}) is equivalent to the one for the SS BH \cite{Katanaev:1997ni}.

For masses close to the Planck mass the $a$ and $B$ dependence of $f$ cannot be neglected anymore and a refined analysis is needed. The more accurate version of (\ref{iso0}) reads
\eq{
T_H=\frac{1}{2\pi}\frac{a}{2(1-a)}(2M_{BH})^{(a-1)/a}\,.
}{iso0better}
The scale parameter $B$ has been fixed such that the next to leading order term of the Killing norm in an inverse ``radius'' expansion is normalized to $-2M_{BH}$. This seemingly harmless normalization has relevant consequences, so probably it is in order to say something about it: while there can be no debate about the leading order term (which can be chosen always equal to $1$) the correct value of the next to leading order term is not completely trivial. The proposed normalization allows to identify $M_{BH}$ at each instant with the mass of an eternal BH. This can be seen most easily by applying the standard subtraction procedure (cf.\ e.g.\ \cite{Liebl:1997ti} or sect.\ 5.1 of \cite{Grumiller:2002nm})
\eq{
M^{\rm reg} = \lim_{r\to\infty} \sqrt{K(r)} \left(\sqrt{K_0}-\sqrt{K(r)}\right)\partial_r X\,,
}{iso44}
where $K(r)$ is the Killing norm of the eternal BH with line element $(ds)^2=2dudr+K(r)(du)^2$ and $K_0$ is the Killing norm of the reference spacetime (in the case of MGS models one takes naturally Minkowski spacetime with $K_0=1$). Applying (\ref{iso44}) to the deformed SS BH (thus treating it at each instant quasi as an eternal BH) yields $M^{\rm reg}=M_{BH}$.

\begin{figure}
\centering
\epsfig{file=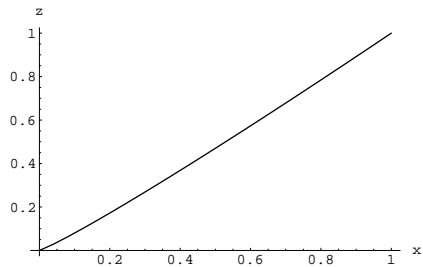,width=.47\linewidth}
\caption{The deviation from a straight line evidently is marginal...}
\label{fig:2.1}
\end{figure}

It is assumed henceforth that $a$ lies in the relevant interval $[0,1)$ and thus $x\in[0,\infty)$. The refined variant of (\ref{iso2}) is given by
\eq{
x=x_i\,z \left(1+\frac{x_i\ln{(x/x_i)}}{\ln{(2M_i)}}\right)^{-1}\,,\quad z=\frac{\ln{(2M)}}{\ln{(2M_i)}}\,.
}{iso2better}
Obviously, for large initial masses the correction terms are negligible (cf.\ fig.\ \ref{fig:2.1} which displays the evolution of a SS BH with a moderate initial mass of $1$ gramme), as long as the ratio of $x/x_i$ does not become too small. However, close to the attractor this is precisely what happens. The BH evaporates down to $M_{\rm Planck}/2$ already at $x=x_i(2M_i)^{-1/x_i}$, which is very close to the attractor. 
\begin{figure}
\centering
\epsfig{file=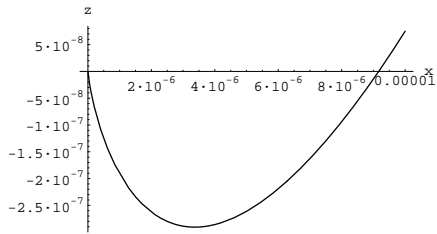,width=.47\linewidth}
\caption{...apart from the final flicker.}
\label{fig:2.2}
\end{figure}

One can try to push the isothermality postulate to the very end. The BH eventually reaches a lower mass limit of 
\eq{
2M_{\rm low} = \exp{\left[-\frac{x_i}{e}(2M_i)^{-1/x_i}\right]}
}{mlower}
at $x_{\rm low}=(2M_i)^{-1/x_i}x_i/e$. For a large SS BH this amounts to approximately half a Planck mass. From that point on isothermal evaporation actually increases the mass slightly until exactly $M_{\rm Planck}/2$ is reached at the attractor solution (cf.\ fig.\ \ref{fig:2.2} which displays the final flicker of a SS BH with initial mass of $1$ gramme). Then only one question remains: is $B$ in (\ref{iso14}) equal or unequal to zero?  A simple analysis proves that $B$ scales with $x$ close to the attractor, so the remnant geometry corresponds to (\ref{iso14}) with $B=0$. It is a regular geometry without any Killing horizon. Its ``mass'' just corresponds to the value of the Killing norm, which is globally normalizable to $\pm 1$ or $0$, depending on its sign. In any case the geometry (\ref{iso14}) is flat for $B=0$. For a positive sign of the last term the coordinates $r$, $u=t-r$ have their usual meaning (radius and retarded time, resp.); for negative sign $u$ corresponds to what usually is denoted by $v=t+r$ (advanced time); for the vanishing case $r$ becomes $v$. It is somewhat amusing that the mass has not really evaporated completely, but has just been converted into a purely geometric entity.

\section{Physical discussion}\label{sec:4}

\subsection{Approximate life time}\label{se:4.e}

In this subsection (and only in this one) some actual meaning will be attached to the coordinate $u$. It will be assumed that the time it takes for the BH to evaporate as measured by the asymptotic observer is actually related to $u$. In the early stages this is obviously justified, because $u\approx\tilde{u}$, where $\tilde{u}$ is the ``physical'' (retarded) time. However, for order of magnitude estimations the difference is negligible because as pointed out before the BH has to radiate away a lot of its mass before it is deformed appreciably.

Applying the $2D$ Stefan-Boltzmann law (\ref{iso0.5}) to the energy loss of an evaporating BH yields
\eq{
\frac{dM}{dt}\propto-(T_H)^2=\rm const.\,,
}{iso6}
The proper life-time estimate then reads
\eq{
t_{\rm evap} \propto \left(M_i\right)^{-1+2/a_i}\,,
}{iso7} 
where the proportionality constant is close to unity. For a SS BH the approximate evaporation time is given by $t_{\rm evap}^{SS}\approx M^3_i$, which agrees with the standard estimation. The reason is that initially the BH for a long time looks like the SS BH---only in the last instances relatively dramatic changes do occur. 

\subsection{The specific heat}

One of the extraordinary properties of the SS BH is its negative specific heat, being quite remarkable from a thermodynamical point of view. Thus it is of interest to check whether this feature changes or not. From (\ref{iso0better}) we obtain
\eq{
c_a:=\left. \frac{\partial M_{BH}}{\partial T_H} \right|_{a=\rm const.} = -\frac{1}{2}\left(4\pi\right)^{a/(a-1)}\left(\frac{1-a}{a}\,T_H\right)^{1/(a-1)}\,.
}{isospecific}
Clearly, the specific heat is negative for the models considered here ($a\in(0,1)$). At $a=1/2$ the SS result $c_a\propto T_H^{-2}$ is recovered. For $a\to 0$ the specific heat vanishes like $a^2/T_H$---thus, as the attractor solution is approached the specific heat vanishes.

It has been advocated in refs.\ \cite{Harms:1992nb} 
that the negative specific heat is not only a sign of instability but also a sign of inconsistency of the canonical approach. Applying a microcanonical analysis a mass loss rate different from the canonical one has been obtained. Thus, also the life-time estimate differs from (\ref{iso7}). It could be interesting to repeat the analysis of the present work for the microcanonical case.

\subsection{Semi-classical issues}\label{se:semi}

Since so many results are available at semi-classical level it is appropriate to discuss at least some of them from the perspective developed in this paper. First of all, the semi-classical derivation of the Hawking flux in the context of $2D$ dilaton gravity (cf.\ e.g.\ \cite{Kummer:1999zy}) 
and the corresponding mass-to-temperature law has been applied to obtain (\ref{iso0better}); in a sense, we have pretended that at each step in the evolution the system can be described by an eternal BH at semi-classical level, but which of the infinitely many eternal BHs is the most proper one changes according to the evolution equation for the deformation parameter (\ref{iso2better}). So each of the intermediate ``eternal'' BHs has to fulfill the semi-classical approximation only for an infinitely small time interval.

We should mention that spherically symmetric quantum deformations of the SS BH have been considered perturbatively already a decade ago by semi-classical means \cite{Kazakov:1994ha}. Also, the fact that dimensional reduction and renormalization need not commute does have a certain impact on semi-classical considerations \cite{Frolov:1999an}. 
Another interesting development is related to the question whether a near extremal (Reissner-Nordstr\"om, Kerr or deSitter) BH can evolve semi-classically towards its extremal limit. There is some evidence that the answer is negative \cite{Barvinsky:2000gf}. 
This would imply that our restriction to non-extremal horizons is justified and singles out the attractor solution as the most likely endstate of BH evaporation in a more general frame. But clearly semi-classical considerations are not sufficient to decide these important questions conclusively. Due to the relation $T_H\propto 1/M_{BH}$ semi-classical methods work very well only for macroscopic SS BHs.

A complementary ansatz has been developed in \cite{Kummer:1997hy,Grumiller:2002dm}: 
instead of integrating out the matter fields on a given background, geometry has been integrated out (exactly) and perturbation theory has been imposed upon the ensuing non-local non-polynomial effective theory, depending solely on external sources and matter degrees of freedom. This approach preserves unitarity and works very well only for microscopic (virtual) BHs \cite{Fischer:2001vz,Kummer:1997hy}. 
Since the involved masses have to be much smaller than $M_{\rm Planck}$ the absence of remnants in these papers is not in contradiction with the results of the present work.

\subsection{The line element}

By a continuous patching of models along $u=\rm const.$ lines a global $2D$ line-element of an evaporating SS BH with initial mass $M_i=M_{\rm ADM}$ can be presented as 
\meq{
(ds)^2=2drdu+(du)^2\Bigg[\theta(u_f-u)\left(1-\frac{2M_{\rm Bondi}(u)}{r^{x(u)}}\right)\\
+\theta(u-u_f)K_{\rm rem}(r,u)\Bigg]\,,
}{iso15}
with $x(u)$ as given by (\ref{iso2better}) (with $z(u)=\ln{(2M_{\rm Bondi}(u))}/\ln{(2M_{ADM})}$) and
\eq{
M_{\rm Bondi}(u)=M_{ADM}-f(u)\theta(u)\,.
}{iso16}
The Bondi mass is assumed to decrease monotonically (apart from the final flicker), i.e.\ $f(u)$ is monotonically increasing and $f(0)=0$ for reasons of continuity. The value $f(u_f)$ is determined by patching to $K_{\rm rem}(r,u_f)$ which encodes the last flicker. The specific form of $f(u)$ depends on the model under consideration. The ``remnant mass'' $M_{\rm rem}$ is of order of $M_{\rm Planck}$, but the further evolution depends on $K_{\rm rem}(r,u)$; it has been argued in sec.\ \ref{se:3.4} that it will look like the Killing norm in (\ref{iso14}) with $B=0$. The explicit solution for $x(u)$ involves the Lambert $W$ function\footnote{For a comprehensive discussion of $W(z)$ see \cite{Corless:1996}. I am grateful to A.~Rebhan for providing this reference.} ($W(z)$ is a multivalued function exhibiting algebraic and logarithmic singularities, but above the limit (\ref{mlower}) only the principal solution for $W(z)$ in $z = W\,e^W$ is needed---for masses below $M_{\rm Planck}/2$ two real branches exist, but a simple continuity argument establishes one of them as the proper one; thus, no branch cut ambiguities arise, except for the point where the limit (\ref{mlower}) is saturated):
\eq{
2M_{ADM} x(u)=\frac{m(u)}{W\left[m(u)\right]}\,,\quad m(u):=2M_{ADM}\ln{(2M_{\rm Bondi}(u))}\,.
}{iso16.5}
For a Bondi mass below $M_{\rm Planck}/2$ a second real branch of $W$ exists; however, continuity of the evolution restricts us to the principal branch. The apparent horizon is located at $r(u)=r_h(u)$ with
\eq{
r_h(u)=\exp{\left[W\left[m(u)\right]\right]}\,.
}{iso16.7}
\begin{figure}
\setlength{\unitlength}{0.01\linewidth}
\begin{picture}(100,30)(0,10)
\put(28,10){\mbox{\resizebox{0.4\linewidth}{0.24\linewidth}{\includegraphics{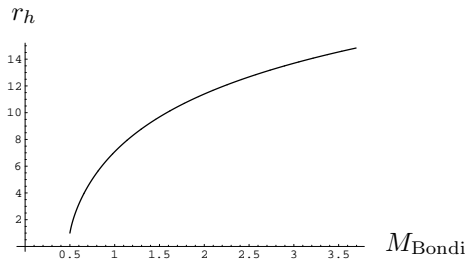}}}}
\put(29,37){\makebox(0,0)[l]{$r_h$}}
\put(70,12){\makebox(0,0)[l]{$M_{\rm Bondi}$}}
\end{picture}
\caption{Apparent horizon as a function of $M_{\rm Bondi}$}
\label{fig:rh}
\end{figure}
It is a monotonically decreasing function with $u$ (see fig.\ \ref{fig:rh}). If we had taken the simpler eq.\ (\ref{iso2}) instead of (\ref{iso2better}) the horizon were located always at $\hat{r}_h=2M_{ADM}$---hence, the Planck scale contribution $df$ in (\ref{iso4}) is responsible for a decreasing apparent horizon. At the lower mass limit (\ref{mlower}) the horizon is located at $r_h^{\rm min}=1/e$. From that moment on, as pointed out before, isothermality would imply an increase of $M_{\rm Bondi}$, somewhat in contradiction to common sense.

\begin{figure}
\centering
\epsfig{file=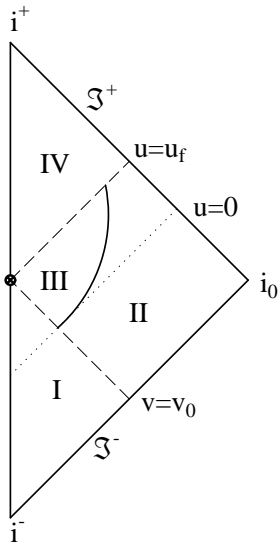,width=0.32\linewidth}
\caption{A sketchy conformal diagram depicting the evaporation of a SS BH}
\label{fig:CP}
\end{figure}

The corresponding Carter-Penrose diagram is depicted in a sketchy manner in fig.\ \ref{fig:CP}. It has to be taken {\em cum grano salis}. For simplicity, region I is taken as fourdimensional Minkowski spacetime. At $v=v_0$ a light-like shock wave forms the SS BH. Region II is the SS region (in this region eventual radiation from the formation process is emitted to $\scri^+$ which has been blithely ignored). At $u=0$ Hawking radiation commences (the light-like line at $u=0$ would be the SS horizon if no evaporation occurred). In region III the continuous deformation towards the attractor solution happens and the apparent horizon decreases by virtue of (\ref{iso16.7}). At $u=u_f$ the minimal mass $M_{\rm low}$ is reached. The last flicker happens approximately at $u=u_f$ (this line has in fact a width of order of Planck length; thus also the full dot at the origin is in fact not a point but a region of order of Planck area). Region IV is the remnant geometry, i.e.\ according to the previous discussion it is Minkowski spacetime.

It is instructive to calculate the $2D$ curvature scalar
\eq{
R(r,u)=-\frac{2a(u)M_{\rm Bondi}(u)}{(1-a(u))^2r^{(2-a(u))/(1-a(u))}}\,.
}{curvature}
For SS the result $R=-4M/r^3$ is recovered. Of course, the fourdimensional curvature scalar vanishes for SS, but not all scalar invariants built from the Weyl tensor do. They are encoded in (\ref{curvature}).

\subsection{Speculations on the last flicker}\label{sec:flicker}

There are three simple ways to deal with the very last stage: either suppose that a small amount of exotic matter really does exist; then the attractor solution is the endpoint of evaporation. However, since one of our postulates was actually positivity of the asymptotic flux this violation, as moderate as it might be, is not a pleasant feature.

Alternatively, it seems tempting to give up isothermality as the lowest mass is reached to avoid exotic matter, i.e.\ to take the line element (\ref{iso15}-\ref{iso16.5}), but assume that (in spite of fig.\ \ref{fig:2.2}) the Bondi mass is decreasing below $M_{\rm low}$ as given by (\ref{mlower}). This has some drastic consequences: $x(u)$ suddenly acquires a non-vanishing imaginary part, $x(u)=\al+i\be$, with $\al,\be\in\mathbb{R}$. Thus, the line element stays real only at the discrete set of points $r=r_0^n$ with $n\in\mathbb{Z}$ and $r_0=\exp{[\pi/\be]}$ (note the duality $r_0\to 1/r_0$). There is a $\mathbb{Z}_2$ grading depending on whether $n$ is even or odd (the quantity $r^{i\be}$ becomes positive or negative, respectively). Eventually the real part goes to zero:
\eq{
2M_{\rm Bondi}(u_0)=\exp{[-\pi/(4M_{ADM})]}<2M_{\rm low}\,\quad\rightarrow\quad x(u_0)=i\be\,.
}{iso42} 
The quantity $\be$ is positive and in fact very small close to the critical point. One can only speculate about the interpretation of the appearance of a discrete structure (or alternatively the appearance of a complex metric). On a superficial level complex and/or discrete structures are expected to arise in a comprehensive theory of quantum gravity; maybe it is most proper to interpret this effect as a phase transition: above $M_{\rm low}$ an adequate description can be achieved by a (smooth) metric, but at $M_{\rm low}$ the system ``condensates'' to some discrete/complex entity. It could be worthwhile to investigate this idea more rigorously\footnote{Since we are already speculating it should be mentioned that the effect of conifold singularities in string theory \cite{Strominger:1995cz,Greene:1995hu} appears to be similar (even the function $x\ln{x}$ plays an essential role in both contexts, cf.\ (\ref{iso2better}) in the current paper and (2.4) in \cite{Strominger:1995cz}). I am grateful to M.~Kreuzer for discussions on that topic.}. Applying the results for the asymptotic limit of the Lambert $W$ function, $W(z\to\infty)=\ln{|z|}+\dots$ where the omitted terms contain multiple logarithms and branch cut ambiguities \cite{Corless:1996}, allows a discussion of geometry in the limit $2M_{\rm Bondi}=:\eps\to 0$:
\eq{
\lim_{\eps\to 0} \left(1-\frac{\eps}{r^{\ln{\eps}/W[2M_{ADM}\ln{\eps}]}}\right) = 1-\lim_{\eps\to 0} \eps^{1-\ln{r}/\ln{|\ln{\eps}|}} = 1
}{iso43}
This produces just the attractor solution.

The third possibility (which is the one favored by the author because of its simplicity and because of the issues discussed below in sect.\ \ref{sec:new}) is to assume that the evaporation process simply stops as the mass drops down to $M_{\rm Planck}/2$. This corresponds to a discontinuous jump from the principal branch to the second real branch of the Lambert $W$ function. Thus, neither a phase transition to a discrete structure nor exotic matter is needed. The ``undershooting'' as depicted in fig.\ \ref{fig:2.2} has been eliminated by cutting the evolution process below $z=0$ and connecting the two (otherwise disconnected) points $x(z=0)$. One can call such a procedure also ``phase transition'' because the Hawking temperature jumps from a finite value to zero. 

Thus, all three approaches yield the attractor solution (\ref{iso14}) with $B=0$ as endpoint---they differ ``only'' in interpretational issues: either one has to allow for (a tiny amount of) exotic matter or for a discrete/complex space-time structure or for a discontinuity as $M_{\rm Planck}/2$ is reached. The second and the third possibility imply the occurrence of a phase transition, because both induce a non-smooth change at a branch point. The order of this transition differs: while for the second scenario continuity has been maintained (and thus one can call it a second order phase transition), the last version implies a first order transition.

It has been shown in sect.\ \ref{se:3.4} that the mass is effectively converted to a purely geometric quantity of the attractor geometry, but energy conservation dictates that something has to happen with the final $M_{\rm Planck}/2$. There are two possibilities:
\begin{enumerate}
\item A remnant in the form of matter concentrated at $r=0$ exists.
\item A thunderbolt emits a total energy of $M_{\rm Planck}/2$ at the end of the evaporation process. 
\end{enumerate} 
Of course, it could be that both possibilities have to be combined.

Suppose for a moment the matter remnant scenario. A point particle in 2+1 dimensions corresponds to a cosmic string in 3+1 dimensions. Then the question arises: Whatever happened to spherical symmetry---how can it be that a spherically symmetric object (the SS BH) evolves into an object with a preferred axis (the cosmic string)? It should be emphasized that the {\em total} mass of the cosmic string is bounded by $M_{\rm Planck}/2$; thus, if it were infinitely long its mass density per unit length went to zero. Possible interactions between cosmic strings and SS BHs have been discussed in the context of ``thorny spheres'' in ref.\ \cite{Frolov:2001uf}; for a physical discussion see section 7.4 of \cite{Frolov:1998} and references therein.
In a more physical setup including also angular momentum it is conceivable that the axis of the remnant string is aligned with the axis of rotation of the macroscopic BH. Moreover, its length is unlikely to be infinite---a typical length scale to be expected would be around Planck length and thus also the mass density will be of order of unity in Planck units. Such a finite size object violates spherical symmetry only moderately. 

\subsection{Dilaton evaporation}\label{sec:new}

As mentioned the variant of a first order phase transition is favored by the author. While revising this paper a new work appeared \cite{Balasin:2003cn} which supports it. When talking about a phase transition one would of course like to know which are the two phases and what separates them---after all, so far we have achieved only a smooth transition towards the attractor geometry, so what can possibly provide the degree of nonsmoothness pivotal to a first order transition? The answer is provided by the dilaton field: while before the evaporation is terminated the dilaton necessarily is non-constant (it can be rearranged to be linear in one of the coordinates by exploiting the gauge degrees of freedom), afterwards it can be constant. Therefore, the phase transition could be from a linear dilaton vacuum to a constant dilaton vacuum. It will be argued in this section that indeed this is the case. Thus, the current approach provides a realization of the semi-classically motivated conjecture \cite{Zaslavsky:1998hp} that constant dilaton vacua might be the final state of BH evaporation.

Most dilaton theories (\ref{vbh:sog}) allow for one or more constant dilaton vacua $X=\rm const.$, where the constant is determined by the zero(s) of the potential\footnote{Note that for the attractor solution $V=0$ and thus the dilaton is unconstrained.} $V(X)$ (cf.\ sect.\ 2.1 of \cite{Grumiller:2003ad}). The corresponding geometry can only be (A)dS, Rindler or Minkowski space. Thus, the only constant dilaton vacuum exhibiting the MGS property is Minkowski space. Because also the attractor solution (\ref{iso14}) with $B=0$ is Minkowski space it seems that the dilaton has evaporated to a constant dilaton vacuum. However, there is an important subtlety (cf.\ sect.\ 3 of \cite{Grumiller:2003ad}): if the attractor solution is patched to such a constant dilaton vacuum along a $u=\rm const.$ line then a matter flux is induced on it, which is remarkable insofar as curvature is continuous (namely zero in our particular case). This provides a mechanism for a shock wave like evaporation of the final $M_{\rm Planck}/2$. 

At this point it should be recalled that the observer is a noninertial one---naive semiclassical extrapolation implies that the boost velocity has to approach the vacuum velocity of light in the limit of vanishing mass; this is nothing but an ultrarelativistic (Aichelburg-Sexl \cite{Aichelburg:1971dh}) limit! Incidentally, the shock-wave induced by such a limit in $2D$ dilaton gravity takes the same form as the shock-wave which arises when patching a constant dilaton vacuum to a non-constant dilaton vacuum \cite{Balasin:2003cn}. Thus, the scenario with the first order phase transition fits nicely together with the boost of the observer implicit in the assumption of isothermality. Actually, exploiting the results of \cite{Balasin:2003cn} one can even provide an educated guess for the total energy emitted in the thunderbolt: only half of the final BH mass will be emitted, because the discontinuous directional derivative of the dilaton, $X^+$ in the notation of \cite{Balasin:2003cn}, jumps from a constant to minus this constant. However, patching to a constant dilaton vacuum implies that it jumps only to zero, i.e., half of the amount as compared to an ultrarelativistic boost---because the height of the jump sets the scale for the energy emitted by the shock wave it seems that only half of the mass of the final state can be emitted in the thunderbolt, the other half will reside in a cold remnant.

Although the first order formulation so far has been avoided it will be considered in this paragraph because there is a nice target space interpretation of the phase transition. A reader who is not familiar with PSMs can skip this paragraph or may consult appendix \ref{app:def} and references therein. There are two basic differences between symplectic manifolds and Poisson manifolds: the Poisson tensor can have a nonvanishing kernel and its rank need not be constant. Indeed, in dilaton gravity there is always a nonvanishing kernel and the corresponding conserved quantity is related to the total energy (essentially the ADM mass, whenever this notion makes sense). But the rank is constant, namely 2, except for eventual isolated points where it vanishes (like the bifurcation 2-sphere in the SS BH). The evaporation to a constant dilaton vacuum corresponds to a transition to a region where the rank of the Poisson tensor vanishes! Thus, before the phase transition the world sheet geometry is defined by symplectic cuts in target space, while afterwards it is defined by a single point in target space.

To conclude, the phase transition is not one between two different geometries (because the line element for the attractor solution is equivalent to the one for the constant dilaton vacuum), but between different behaviors of the dilaton field. It can be understood in terms of an ultrarelativistic boost of the observer. A similar mechanism could be of relevance for scalar-tensor theories in $D=4$; there, a constant dilaton vacuum simply corresponds to a phase where the effective Newton constant does not vary.

\subsection{The information paradox}\label{se:4.4}

It is difficult to avoid addressing the information loss puzzle (cf.\ e.g.\ \cite{Preskill:1992tc}) 
when discussing the long time evolution of BHs. The main two alternatives are 1.\ the information is lost, as proposed by Hawking \cite{Hawking:1976ra} 
and 2.\ some mechanism exists that prevents the information loss. Most people with a field (or string) theory background favor the second variant. In the following it will be discussed very briefly which of the proposed solutions fits into the current approach.

The quantum hair solution does not fit into the current formalism since the formation process has been neglected. Also the possibility that the information is emitted together with the Hawking radiation is not applicable here, at least not without specifying the matter content of the theory; neither is the baby universe scenario, at least not in an obvious way. Since a remnant geometry has been predicted there remain two natural solutions: either the information is retained in the remnant or it emerges at the end of the evaporation process. Due to the fact that the proposed remnant geometry is flat it is hard to see where any information could be stored (apart from the scenario with a cosmic string remnant), so the only possibility seems to be that the information is released during the final stages. Indeed, this is what the Carter-Penrose diagram fig.\ \ref{fig:CP} seems to suggest, as well as the discussion in sect.\ \ref{sec:new}. It will be studied in more detail below.

The information paradox manifests itself typically in a pathological feature of Carter-Penrose diagrams related to BH evaporation (cf.\ e.g.\ fig.\ 1 of \cite{Giddings:2004ud} for the classical example; in that reference it is argued in favor of a vioation of locality). Thus, it is worthwhile to study the present Carter-Penrose diagram fig.\ \ref{fig:CP} and to check what, if any, pathological features are encoded there. To this end the consideration of test particles will be useful. Note that in the language of the S-matrix approach to quantum black holes \cite{Stephens:1994an} 
the ``hard matter'' components implicitly have been taken into account by our assumptions. ``Hard'' refers to the fact that its backreaction upon geometry cannot be neglected, as opposed to ``soft'' matter which can be treated like test particles. It will be assumed for simplicity that the test particles are massless.

Let us start at some value of advanced time $v=v_{\rm in}$ at $\scri^-$ and follow the light rays until they are reflected at the origin and eventually emitted to a point at $\scri^+$ corresponding to a certain value of the retarded time $u=u_{\rm out}$. If $v_{\rm in}$ is sufficiently close to $i^-$ then $u_{\rm out}<0$, i.e.\ the outgoing ray will pass solely through region I, encounter the infalling matter flux and finally pass through II. Similary, for values of $v_{\rm in}$ sufficiently close to $i^0$ the lightray will pass through regions II and III, encounter the outgoing shock wave and end up in region IV where eventually it will be emitted to $\scri^+$ close to $i^+$. Actually, this is true for all values $v_{\rm in}>v_0$. Of course, rays starting closer to $v_0$ will additionally pass the apparent horizon, but so far no pathological features were found. At $v=v_0$ there is the problem of the potential singularity encountered at the origin. Recently it has been proposed to impose a boundary condition at the singularity \cite{Horowitz:2003he} (cf.\ \cite{Gottesman:2003up} for a critical comment) as a possible path to a solution to information loss; in the present context it is not quite clear whether the singularity exists at all as it lies in the very region we could only speculate about. So let us diregard it and focus on the region at $\scri^-$ that has not been covered so far: $v_{\rm in}<v_0$ but still sufficiently far away from $i^-$ such that light rays are not scattered to region II.

Such a test particle will trespass the line $u=0$ and be reflected at the origin still in region I. However, it crosses the infalling matter shock wave between $u=0$ and $u=u_f$ thus entering region III. It will follow on a curved path more or less parallel to the apparent horizon and then pass the outgoing shock wave. In region IV it moves first again towards the origin and after the reflection it finally reaches $\scri^+$. Such test particles are remarkable for three reasons: 1.~they are reflected twice at the origin, 2.~they cross infalling {\em and} outgoing shock wave, 3.~for each test particle travelling in this way there exists exactly one ``partner particle'' emitted at $v_{\rm in}>v_0$. The ``partner particle'' will end up at the same point at $\scri^+$. Thus, tracing back the light rays from $\scri^+$ there exists a region between $u=u_f$ and some larger (but finite) value of $u$ where the light ray splits into two as soon as the outgoing shock wave is encountered from above: one directly going back to $\scri^-$ and the other one entering region III. In this sense, the outgoing shock wave causes birefringence. The line $v=v_0$ acts like a mirror regarding the two ``partner particles'': if one of them is close to (far from) $v_0$ so is the other. 

On top of that, the {\em later} the test particle is emitted at $\scri^-$ in region I (i.e.\ the closer $v_{\rm in}$ is to $v_0$) the {\em earlier} it arrives at $\scri^+$ in region IV (i.e.\ the closer it is to $u=u_f$). Thus, causality is violated (or rather: inverted), but this violation is not noticed at $\scri^+$ until the BH has evaporated completely!

This is the pathological feature we were looking for (or hoping to avoid): there is something very strange about $\scri^+$: while the regions $u<0$ and close to $i^+$ are unproblematic, there is no way to send information to the region $u_f>u>0$ by means of test particles---this coincides with the region dominated by Hawking radiation caused by ``hard matter''. Moreover, there exists a region $u_e>u>u_f$ with some finite $u_e$ where information may arrive in two ways: either directly from $\scri^-$ with only one reflection or with two reflections and inverted causality, as described in the previous paragraphs. It is worthwhile to study these features in more detail, but such a study would lead beyond the scope of this paper.

\section{Relaxing the conditions and conclusions}\label{sec:5}

\paragraph{Relaxations within $\boldsymbol{2D}$ dilaton gravity} There are three generic possibilities regarding the first assumption: first, one can assume the complete absence of deformations with the usual ensuing problems; second, one can allow for consistent deformations only as in the present approach (with the important caveat that only deformations of the spherically reduced theory have been considered); third, one can allow for arbitrary deformations which enlarges the ``phase space'' of possible models but at the same time reduces the predictability. It is sound to start with the second possibility until some fundamental theory dictates the precise form of the allowed deformations. 

Relaxing the second assumption is possible in principle---e.g.\ a BH immersed in (A)dS or Rindler spacetime can be considered along these lines. However, the fact that asymptotics is not changed essentially during the evaporation process should be maintained, because otherwise the notion of a long-time asymptotic observer will be challenged. For the third assumption similar considerations apply: one can allow e.g.\ for 2 Killing horizons and thus is able to discuss the Reissner-Nordstr{\"o}m case. Such a generalization is not as harmless as it might seem at first glance: once 2 or more horizons are allowed extremal ones may emerge during the evaporation process. Since they do not produce Hawking radiation extremal geometries are additional candidates for possible endpoints of the evolution. Because stringy methods are capable to describe (nearly) extremal BHs very well (cf.\ e.g.\ \cite{Horowitz:1996qd}) 
one could imagine that the non-extremal evolution is described as discussed in the current work and from the extremal point on string theory takes over. However, there is classical and semi-classical evidence that nearly extremal BHs do not evolve towards the extremal limit, which can be taken as a justification of our neglection of extremal horizons (cf.\ sect.\ \ref{se:semi}). 

Relaxing the crucial fourth assumption does not seem plausible from a physical point of view---either the asymptotic flux would have to be negative at some point (``anti-evaporation'') or it would have to grow without bound. 

Obviously it is desirable to generalize from the $a-b$ family to a more general class (or the most general class) of possible models. Some steps in that direction are presented in appendix \ref{app:A}. As long as no extremal horizon is present evidence points again towards the previously discussed remnant geometry as the endstate of evaporation.

\paragraph{Beyond $\boldsymbol{2D}$ dilaton gravity} The restriction to PSMs\footnote{A generalization to graded PSMs in order to describe dilaton supergravity models should be straightforward, cf.\ \cite{Strobl:1999zz}.}, 
even though it allows for phenomenologically relevant models like the SS BH or generalizations thereof (with charge and/or cosmological constant), could be overcome by leaving the comfortable realm of $2D$. Due to the rigidity result \cite{Barnich:1995ap} mentioned in footnote \ref{fn:7} a natural framework would be a scalar-tensor theory a la Jordan-Brans-Dicke \cite{Fierz:1956} 
or its more recent incarnation as quintessence \cite{Wetterich:1988fm}. 
A brief discussion of scalar-tensor theories in the framework of $2D$ models can be found in \cite{Grumiller:2000wt}. It should be possible to apply the methods of this work to the dimensionally reduced models treated in that reference. 

To be a bit more concrete: an action of a $D$-dimensional scalar tensor theory,
\begin{equation}
  \label{eq:stt1}
  \int_{M_D} d^Dx\sqrt{-g}\left[XR-U(X)(\nabla X)^2+2V(X)\right]\,,
\end{equation}
contains two arbitrary functions $U,V$ of the dilaton $X$ which usually are adjusted either by hand or by some guiding principle from a more fundamental theory. Tinkering with the assumptions, it should be possible to generalize the present work from $D=2$ to arbitrary $D$, in particular to $D=4$. If again the phenomenon of ``dilaton evaporation'' to a constant dilaton vacuum occurs, the final state will be a solution of ordinary vacuum Einstein theory with a cosmological constant $\Lambda$, the value of which depends on the function $V(X)$ of the attractor solution: $\Lambda\propto V(X_{CDV})/X_{CDV}$, where $X_{CDV}$ is a solution of the equation $2V(X_{CDV})=X_{CDV}V'(X_{CDV})$. The existence of such a solution poses nontrivial constraints on the dilaton potential $V$.

\paragraph{What are the experimental consequences?} Since for each SS BH (independent of its original mass, as long as it was larger than $M_{\rm Planck}/2$) a cosmic string with a total mass of maximal $M_{\rm Planck}/2$ remains this puts an upper bound on the total amount of primordial BHs. Actually, this bound is well-known \cite{Green:1997sz} because several models predict a relic with a mass of order of $M_{\rm Planck}$ \cite{Bowick:1988xh}. 
Apart from their total mass cosmic strings contribute to density fluctuations which should be visible in the observed cosmic microwave background anisotropies (cf.\ e.g. \cite{Bouchet:2000hd} for an analysis of the BOOMERanG data). Finally, the thunderbolt with a total energy of maximal $M_{\rm Planck}/2$ should lead to observable consequences, possibly in the form of $\ga$-ray bursts. The sum of the energy of the thunderbolt and the mass of the cosmic string remnant is exactly $M_{\rm Planck}/2$ provided energy conservation holds. However, none of these consequences provides a ``smoking gun signature''. Moreover, as mentioned before, it is not unlikely that eventually non-spherical modes will become relevant during the evaporation process, which further limits the phenomenological relevance of the present work. 

\paragraph{Brief summary} Let us finally repeat the lines of reasoning underpinning the approach presented in this paper: it is generally expected that during the final stages of BH evaporation some drastic changes will occur, since the semiclassical theory extrapolated to the very end predicts its own failure. In order to discuss the issue of long time BH evaporation in more detail either quantum gravity must be applied or some trick must be used. It has been proposed to follow the latter route by assuming an isothermal observer at $i^0$, i.e.\ the asymptotic Hawking flux (and hence the Hawking temperature) as measured by this observer does not change during the whole evaporation process. It can be replaced by a much weaker condition, namely boundedness and positivity of the asymptotic Hawking flux. As a consequence---together with ``natural'' assumptions regarding the asymptotics, the causal structure, the nature of allowed quantum deformations of gravitational symmetries and some simplifying technical assumption further restricting the allowed deformations---an evolution of the SS BH towards a remnant geometry (\ref{iso14}) with $B=0$ has been predicted. 
We speculated about the final flicker, argued in favor of a first order phase transition to a constant dilaton vacuum, discussed briefly implications for the information paradox and concluded with possible generalizations of our postulates.

If nothing else, the present approach shows that quantum induced deformations of the gauge symmetries of gravity can play a pivotal role for the understanding of the long time behavior of black holes.

\section*{Acknowledgement}
 
This work has been supported by projects P-14650-TPH and J-2330-N08 of the Austrian Science Foundation (FWF). I would like to thank my long-time collaborators on $2D$ gravity, W.~Kummer and D.~Vassilevich, for numerous stimulating discussions in the past, present and future and for a careful reading of the manuscript. Moreover, I am grateful to H.~Balasin for his interest, valuable comments and his important role as {\em advocatus diaboli}. I render special thanks to T.~Strobl for bringing ref.\ \cite{Izawa:1999ib} to my attention and to F.~Brandt for e-mail correspondence on consistent deformations. I have profited from conversations and e-mail exchanges with D.J.~Schwarz on cosmological topics. I thank A.~Barrau, A.~Bonanno, D.~Easson, B.~Harms and S.~Nojiri for pointing out some references and/or for correspondence. Several comments by the anonymous referee led to a considerable improvement of the presentation. Finally, I am grateful to V.~Frolov for discussions on cosmic strings and SS BHs and to O.~Zaslavskii for a useful pedagogical suggestion during the workshop ``Gravity in two dimensions'' at the International Erwin-Schr\"odinger Institute.

\begin{appendix}

\section{Deformations of dilaton gravity}\label{app:def}

This appendix is meant to be self-contained. Its purpose is to review concisely some features of ordinary and deformed dilaton gravity which are (at least implicitly) relevant for the rest of this paper. 

To this end (only in this appendix) the first order version of (\ref{vbh:sog}) \cite{Schaller:1994es,Strobl:1999wv}
\begin{equation}
L^{(1)}=\int_{\mathcal{M}_2} \left[
X_a (D\wedge e)^a +Xd\wedge\omega +\epsilon \mathcal{V} (X_aX^a, X) \right]\,,
\label{eq:dvva1}
\end{equation}
will be employed where $X$ is the dilaton field, $e^a$ is the zweibein one-form, $\epsilon$ is the volume two-form. The one-form $\omega$ represents the  spin-connection $\om^a{}_b=\eps^a{}_b\om$ with $\eps_{ab}$ being the totally antisymmetric Levi-Civit{\'a} symbol. The action (\ref{eq:dvva1})  depends on two auxiliary fields $X^a$. It is a special case of a PSM \cite{Ikeda:1993aj,Schaller:1994es} 
\eq{
L^{PSM} = \int_{\mathcal{M}_2} \left[dX^i \wedge A_i + \frac{1}{2} \biv^{ij} A_j \wedge A_i\right]
}{eq:ka2}
with the Poisson-tensor ($f$ and $\cas$ are arbitrary functions of $X^i$) 
\eq{
\biv^{ij} = \eps^{ijk} f \cas_{,k} := \eps^{ijk} f \frac{\partial\cas}{\partial X^k}
}{eq:ka1}
which fulfills the (generalized) Jacobi identity
\eq{ 
P^{in}P_{,n}^{jk}+{\rm cycl.}(i,j,k) = 0\,,
}{eq:jac}
and a three dimensional target space the coordinates of which are $X,X^a$. In light-cone coordinates ($\eta_{+-}=1=\eta_{-+}$, $\eta_{++}=0=\eta_{--}$) the first (``torsion'') term of (\ref{eq:dvva1}) is given by
\begin{equation}
X_a(D\wedge e)^a = \eta_{ab}X^b(D\wedge e)^a =X^+(d-\omega)\wedge e^- +
X^-(d+\omega)\wedge e^+\,.\label{dvvXDe}
\end{equation}
The function $\mathcal{V}$ is an arbitrary potential depending solely on Lorentz invariant combinations of the target space coordinates, namely $X$ and $X^+X^-$. To match the first order with the second order formulation one has to choose $\mathcal{V}=U(X)X^+X^-+V(X)$ with the same functions $U,V$ as in  (\ref{vbh:sog}). To match the PSM with the first order version one has to choose $f=1/I(X)$, $\cas=I(X)X^+X^-+w(X)$ with
\eq{
I(X):=\exp{\int^XU(y)dy}\,,\quad w(X):=\int^XI(y)V(y)dy\,.
}{def:1}

Variation of the fields yields the equations of motion
\begin{align}
dX^i + P^{ij}A_j &= 0\,, \label{eq:psmpvn6} \\
d\wedge A_i - \frac{1}{2}P_{,i}{}^{jk} A_k \wedge A_j &= 0\,. \label{eq:psmpvn7} 
\end{align}
Under the symmetry variation (note that $\eps=\eps(x^\mu,X^i)$, where
$x^\mu$ are the world-sheet coordinates and $X^i$ the target space 
coordinates)
\eq{
\de_\eps X^i = P^{ij}\eps_j\,, \hspace{0.5cm} \de_\eps A_j = - d\eps_j -
P_{,j}{}^{kn} \eps_n A_k\,,
}{eq:psmpvn10}
the action (\ref{eq:ka2}) transforms into a total divergence
\eq{
\de_\eps L = \int_{\partial \mathcal{M}_2} dX^i \eps_i\,.
}{eq:psmpvn11}
The commutator of two symmetry variations closes only on-shell in general:
\begin{align}
\left[\de_{\eps_1},\de_{\eps_2}\right]X^i &= \de_{\eps_3} X^i\,, 
\label{eq:psmpvn12} \\
\left[\de_{\eps_1},\de_{\eps_2}\right]A_i &= \de_{\eps_3} A_i + \left(dX^j+
P^{jk}A_k\right)P_{,ji}{}^{nm} \eps_{1\,m} \eps_{2\,n}\,, 
\label{eq:psmpvn13} 
\end{align}
with
\eq{
\eps_{3\,i} = P_{,i}{}^{jk} \eps_{1\,k} \eps_{2\,j} +
P^{jk} \left[\eps_k,\left(\eps_{,j}\right)_i\right]_{12}\,,
}{eq:psmpvn14}
and $[f(\eps),g(\eps)]_{12}:=f(\eps_1)g(\eps_2)-f(\eps_2)g(\eps_1)$. 

The introduction of the Schouten-Nijenhuis bracket \cite{Schouten:1954} 
\eq{
\left\{X^i,X^j\right\}=P^{ij}\,,\quad\left\{X^i,\bullet\right\}=P^{ij}\partial_j (\bullet)\,,
}{def:2}
allows to reexpress e.g.\ the conservation law as $\{X^i,\cas\}=0$, $\forall i$. According to Izawa \cite{Izawa:1999ib}, consistent deformations in the sense of Barnich and Henneaux are those which leave the bracket structure (\ref{def:2}) essentially intact, i.e.\ the Poisson tensor may change its functional dependence on the fields $X^i$, but its dimension (and the fact that one stays within the realm of PSMs) remains fixed. 

Since Izawa's result is of some importance for the present approach his proof will be sketched briefly: starting point is $2D$ BF theory (the conventions of \cite{Izawa:1999ib} have been translated accordingly)
\eq{
S_0 = \int_{\mathcal{M}_2} dX^i\wedge A_i\,,
}{eq:iza1}
which is just a very trivial special case of a PSM (\ref{eq:ka2}) with vanishing Poisson tensor. Then, the minimal solution $\bar{S}$ to the classical master equation and the generator of BRST symmetry $s$ are constructed. A deformation $L_1$ of the Lagrangian must obey \cite{Barnich:1993vg}
\eq{
sL_1+da_1=0\,.
}{eq:iza2} 
This, together with its descent equations $sa_1+da_0=0$, $sa_0=0$, allows to construct the most general consistent $L_1$. It is given by eq.\ (14) of \cite{Izawa:1999ib}. The deformed action
\eq{
S=\bar{S}+\int_{\mathcal{M}_2} L_1 = S_0 + \frac{1}{2}\int_{\mathcal{M}_2} P^{ij} A_j\wedge A_i + \rm antifield\, terms\,,
}{eq:iza3}
satisfies the classical master equation and reduces to (\ref{eq:ka2}) for vanishing antifields. Actually, it even satisfies the quantum master equation as there is no contribution from renormalization \cite{Cattaneo:1999fm}. No further deformations occur because $P^{ij}$ fulfills the (generalized) Jacobi-identity (\ref{eq:jac}), i.e.\ the first order deformation of abelian BF theory, which is nothing but a PSM, provides already the most general consistent deformation. Thus, also the most general consistent deformation of a PSM is just another PSM (with the same number of target space coordinates).

It is important to notice that dilaton gravity, if formulated as a PSM, requires additional structure, namely a correspondence between the gauge fields and the line element. The suggestive notation $A_i=(\om,e^-,e^+)$ together with $g=2\eta_{+-}e^+e^-$ apparently fixes this structure in the undeformed case, but there is no a priori relation between the three gauge fields and the line element if deformed PSMs are considered. This subtle and important issue is addressed in detail in \cite{Strobl:2003kb}.

\section{The $\boldsymbol{U-V}$ family}\label{app:A}

For technical reasons the main part of the paper concentrated on the $a-b$ family. In this appendix the results will be generalized to the $U-V$ family, i.e.\ to models with a Lagrangian (\ref{vbh:sog}) with arbitrary potentials $U(X)$ and $V(X)$ which fulfill the MGS condition and which have at most one horizon. 

All local and global classical solutions of such models have been discussed extensively in a series of papers by T.~Kl\"osch and T.~Strobl \cite{Klosch:1996fi}. 
For a first orientation on classical, semi-classical and quantum dilaton gravity in $2D$ the review ref.\ \cite{Grumiller:2002nm} may be consulted.
 
In the following the notation and results of \cite{Grumiller:2002dm} will be used; for sake of self-containment we recall two definitions: the ``integrating factor'' $I(X)$ and the conformally invariant combination $w(X)$ as defined in (\ref{def:1}). The MGS property reads $w\propto 1/I$. Provided that the equation $I=c$ has exactly one solution for positive values of $c$ and no solution for $c\leq 0$ the existence of at most one horizon is guaranteed. The constant $c$ scales like $1/M_{BH}$. The Hawking temperature reads $T_H\propto U(X_h)M_{BH}$ where $X_h$ is the solution of $I(X_h)=1/M_{BH}$. It is convenient to introduce the inverse function of the integrating factor, $I^\ast\circ I=\rm id$, because $X_h=I^\ast(1/M_{BH})$. With these preliminaries the Hawking temperature for MGS models of the $U-V$ family with one horizon reads
\eq{
T_H\propto \frac{M_{BH}^2}{\left.(I^\ast)'\right|_{1/M_{BH}}}\,.
}{app:1}

Now we would like to address the issue of boundedness of $T_H$. Suppose that $I^\ast$ is a given function of $X$ (from this one can deduce all other quantities). First, it will be assumed that $I^\ast$ has a Laurent series expansion of the form
\eq{
I^\ast = \sum_{n=-\infty}^Na_nX^n\,,\quad N\in\mathbb{Z}\,.
}{app:2}
If $N=0$ then the leading term drops out after differentiation, so in this case the value for $N$ relevant in all subsequent formulas is given by the next to leading order term. For small values of the BH mass the Hawking temperature is dominated by the highest power $N$ in (\ref{app:2}) (and hence the discussion is essentially reduced to the $a-b$ family),
\eq{
T_H(M_{BH}\to 0)\approx a(M_{BH})^{N+1}\,,\quad a\in\mathbb{R^+}\,.
}{app:3}
Thus, only for $N\ge -1$ (but $N\neq 0$, because this term drops out after differentiation) the temperature remains bounded for all times. 
Relevant examples/counter examples are the CGHS model with $N=-1$ and spherically reduced gravity from $D=4$ with $N=-2$, respectively. The attractor solution does not fit into this scheme because $I=\rm const.$ does not allow for an inverse (but this model is rather trivial anyway).

If there is a nonperturbative prefactor in front of the sum in (\ref{app:2}) which is regular at $M_{BH}=0$ then again the highest term dominates and the same bound $N\ge -1$ is recovered. If there is a different (but at $M_{BH}=0$ regular) nonperturbative prefactor for each term then still the same conclusion holds. Thus, the bound $N\ge -1$ is rather general. If the prefactor is singular a different bound may be obtained (e.g.\ $N>-1$ for a logarithmic prefactor).

If the sum in (\ref{app:2}) does not terminate (i.e.\ $N\to+\infty$) then an essential singularity at $M_{BH}=0$ is encountered. In order to decide whether $T_H$ remains bounded again the limit $M_{BH}\to +0$ has to be considered. In cases where the Laurent series can be resummed this is straightforward (e.g.\ for something like $\exp{[-1/M_{BH}]}$).

In principle one can also extract an MGS theory for a given temperature law (if, for instance, the temperature law can be motivated by other means). The only MGS model with $T_H=\rm const.$ turns out to be the CGHS. The only MGS model with $T_H=0$ is the (trivial) attractor solution. Thus, if Hawking radiation is assumed to stop at a certain point the attractor solution is the unambiguous remnant (unless, of course, one allows for extremal horizons).

A generalization of these considerations to non $U-V$ theories (like e.g.\ the class of models discussed in \cite{Grumiller:2002md}) is possible in principle, but somewhat tedious in detail.

\section{A toy model from classical mechanics}\label{sec:cm}

Suppose a dynamical system which can be described by a Lagrangian action depending on two fields, one of them being called ``geometrical'' ($q$) and the other one ``matter'' ($Q$). Suppose for simplicity that we are in 1+1 dimensions and that the geometrical part alone can be cast into a 0+1 dimensional form (assuming staticity), but that we do not know how to solve the coupled system. Instead, deformations of the geometrical subsystem will be studied and assumptions will be made which finally allow to obtain an attractor solution without having to specify the matter part.

To keep it basic, let the undeformed geometrical action be a harmonic oscillator
\begin{equation}
  \label{eq:rev1}
  S_0 = \frac12\int\left(\dot{q}^2-q^2\right)dr\,,
\end{equation}
where $\dot{q}:=dq/dr$ and $r$ is a quantity to be interpreted as ``radius''. There is no gauge degree of freedom and one physical degree of freedom. In analogy to appendix \ref{app:def} we want to find all consistent deformations in the technical sense of Barnich and Henneaux \cite{Barnich:1993vg}---i.e.\ the number of gauge degrees of freedom is not changed as well as the number of physical degrees of freedom, but the action may change as well as the gauge symmetries. Often a perturbative analysis in terms of the deformation parameter is employed, but in the present paper the model is simple enough to go beyond perturbation theory. For the toy example deformations are rather trivial,\footnote{First, an antifield $q^\ast$ is defined with antibracket $(q,q^\ast)=1$; then the BRST charge is constructed, $s=\ddot{q}\de/\de q^\ast$; finally, the deformation equation $sL^{(1)}+da_0=0$ together with its decent equation $sa_0=0$ have to be solved, where $L^{(1)}$ is the deformed Lagrangian. These methods \cite{Barnich:1993vg} (cf.\ also \cite{Henneaux:1998bm} and references therein) are being used in appendix \ref{app:def}, but appear to be a slight overkill for the toy example.} 
\begin{equation}
  \label{eq:rev2}
  S_1 = S_0 - \int V(q,\dot{q},\ddot{q},\dots;\tilde{\eps})dr\,,
\end{equation}
where $\tilde{\eps}$ is the deformation parameter and $V(q,\dot{q},\ddot{q},\dots;0)=0$. We will simplify the issue further by requiring independence of $\dot{q}$ or higher derivatives and impose linearity in $q^2$ and in the deformation parameters (these requirements correspond to some of the ``physical'' assumptions that are imposed on the system and of course their validity should be justified; however, as this is merely a toy example this is omitted). Thus, we have a 1-parameter family of deformed geometrical actions at our disposal,
\begin{equation}
  \label{eq:rev3}
  S_\eps = \frac12\int\left(\dot{q}^2-\eps q^2\right)\,.
\end{equation}
Initially, $\eps=1$. Let us impose the ``asymptotic'' condition $q(0)=0$ and assume that the observer is located in the ``asymptotic region'' $r=0$. Then, the general solution of the equations of motion to be derived from (\ref{eq:rev3}) is
\begin{equation}
  \label{eq:rev4}
  q(r;\eps\neq 0) = A\sin{(\sqrt{\eps}r)}\,,\quad q(r;\eps=0)=Br\,,\quad A,B\in\mathbb{R}\,.
\end{equation}
The asymptotic observer patches together these solutions along $t=\rm const.$ lines, where $t$ is to be interpreted as ``time''. Now let me define a ``physical observable'' which I would like to call ``toy model Hawking temperature'':
\begin{equation}
  \label{eq:rev5}
  T_H:=\dot{q}(0)=A\sqrt{\eps}\,.
\end{equation}
Assume that the system can be reparametrized in such a way, that in the full dynamical system there is no time-dependence of $T_H$. Provided $A$ is positive and strictly monotonically increasing, $\eps$ has to be strictly monotonically decreasing. In the limiting case $A\to\infty$ the deformation parameter $\eps$ must vanish. Thus, the existence of an attractor solution
\begin{equation}
  \label{eq:rev7}
  q_{\rm att}(r)=Br\,,\quad B\in\mathbb{R}\,,
\end{equation}
has been established. However, nothing can be said about the speed of the transition to the attractor solution and of course its soundness depends on the validity of the assumptions made to derive it. Finally, it has to be checked whether the constant $B$ in (\ref{eq:rev7}) has to scale to zero or not.

Within this toy model it is even possible to provide a matter action explicitly:
\begin{equation}
  \label{eq:rev17}
  S_{(m)} = \frac12 \int \left(q^2 Q + (dQ/dt)^2 - U(Q)\right)drdt
\end{equation}
If $U(Q)$ is such that the solution of $d^2Q/dt^2 + U'(Q) + {\mathcal O}(q^2)=0$, $Q(t)$, coincides with $(1-\eps)$ for all $t$, then the patching procedure above reproduces the exact dynamics of the whole system close to the asymptotics $r=0$ (which is the region relevant for the observer), because the equation of motion $\ddot{q}+q(1-Q)=0$ from (\ref{eq:rev1}) and (\ref{eq:rev17}) becomes equivalent to the one from (\ref{eq:rev3}). However, it is emphasized that for the BH evaporation discussed in the rest of this paper no attempt is made to construct a corresponding matter action.

\end{appendix}




\providecommand{\href}[2]{#2}\begingroup\raggedright\endgroup


\end{document}